**Viral air load due to sedimenting and evaporating droplets produced by speaking**

Roland R. Netz, Physics Department, Freie Universität Berlin, 14195 Berlin, Germany

*Abstract*: The effect of evaporation on droplet sedimentation times is crucial for estimating the risk of infection from virus-containing airborne droplets. For droplet radii in the range 100 nm < R < 60 µm, evaporation can be described in the stagnant-flow approximation and is diffusion limited. Analytical equations are presented for the droplet evaporation rate, the time-dependent droplet size and the sedimentation time, including the significant effect of evaporation cooling. Evaporation makes the time for large droplets to sediment much longer and thus significantly increases the viral air load. Using recent estimates for SARS-CoV-2 concentrations in sputum and droplet production rates while speaking, a single infected person that constantly speaks without a mouth cover produces a total air load of more than $10^4$ virions. In a mid-size closed room, this leads to a viral inhalation frequency of at least 2.5 per minute. Low relative humidity, as encountered inside buildings in winter and in airliners, speeds evaporation and thus keeps initially larger droplets suspended in air.

**1. Introduction**

In the context of airborne viral infection pathways, the sedimentation properties of liquid droplets that contain non-volatile solutes and are subject to gravitational force, evaporation and evaporation cooling, are crucial. Partial aspects of this problem have been treated in previous experimental and theoretical works (1; 2; 3; 4; 5; 6; 7; 8; 9; 10) (11; 12). For estimates of the infection risk from airborne virus-containing droplets, the relevant droplet radii are less than 50 µm, because only those droplets stay floating in air for sufficiently long time. Duguid studied droplet sizes produced by humans sneezing, coughing and speaking from a microscopic analysis of marks left on slides and found droplet radii between 1 and 500 µm (3). In fact, 95% of all particles had radii below 50 µm, and most final droplet radii were around 5 µm. Later studies basically confirmed these results and showed that in addition many droplets are produced in the sub-micron range during coughing and speaking (13; 14; 15; 16; 17; 18). In one study a multimodal droplet size distribution was found and rationalized in terms of distinct physiological droplet production mechanisms (19). It was shown that the number of droplets produced while speaking depends among other factors on the voice loudness (20) and that droplet production while exhaling is the product of complex fluid fragmentation processes (21). Recently, a much more sensitive method, time-resolved laser-light scattering, showed that far more droplets are produced than could be detected previously (22; 23), which demonstrates that the measured droplet radius distribution depends on the size sensitivity of the measurement technique used and also on the time droplets spend in air before measurement. In the present work, evaporation effects for droplets with radii in the range from nm to a few hundred µm are considered, which is the range potentially relevant for the airborne route of virus infection (24; 25; 26; 27). The calculations include the interplay of all relevant physical effects: i) the maximal evaporation reaction rate at the droplet surface as a function of relative humidity, ii) concentration-boundary as well as flow-boundary layers, iii) droplet cooling due to the large evaporation enthalpy of water, and iv) the water vapor pressure reduction due to the presence of non-volatile solutes (including virions) in the droplet. Analytical expressions for the evaporation rate, the time-dependent droplet radius and the sedimentation time are derived in all relevant radius regimes and relative humidities and estimates for the viral air load from speaking are derived, from which the virion inhalation frequency in closed rooms including air exchange due to ventilation is calculated.

Evaporation effects are typically treated on the level of the diffusion equation in the stagnant air approximation, i.e. neglecting the flow field around the droplet, and in the diffusion-limited evaporation regime. As shown here, this approximation is accurate for droplet radii in the range 100 nm < $R$ < 60 μm, where evaporation cooling is important and reduces the droplet surface temperature by about 9 Kelvin at a relative humidity (RH) of 0.5, which significantly slows down evaporation. For radii larger than 60 μm, the air flow around the droplet speeds the evaporation process and at the same time becomes non-Stokesian due to non-linear hydrodynamics effects, which is treated analytically by double-boundary-layer theory including concentration and flow boundary layers. For radii smaller than 100 nm, the evaporation at the droplet-air interface becomes reaction-rate limited. For these small droplets, the evaporation rate is not limited by the speed with which water molecules diffuse away from the droplet surface, but rather by the rate at which water evaporates from the liquid surface.

In the presence of evaporation, the sedimentation time is determined by the final dried-out droplet radius, which depends on relative humidity and the initial solute concentration. Evaporation makes large droplets remain in air much longer and thus significantly increases the airborne viral load. Using recent estimates of the *SARS-CoV-2* concentration in sputum (28) and droplet production rates while speaking (22) (23), a single person that is infected and speaks constantly is predicted to produce an airborne viral air load in the steady state of more than $10^4$ virions. In a mid-size closed room, this will result in a virion inhalation frequency by a passive bystander of at least 2.5 per minute, which is only mildly reduced by air-exchange rates in the typical range of up to about $20\ min^{-1}$. These numbers clearly demonstrate the possible significance of air-borne viral infection pathways.

## 2. Droplet sedimentation and diffusion without evaporation

It is useful to first recapitulate a few well-known basic equations in the absence of droplet evaporation. By balancing the Stokes friction with the gravitational force, proportional to the acceleration $g$, that acts on a droplet with radius $R$ and mass density $\rho$, the mean sedimentation time (see Appendix A) is

$$\tau_{sed} = \frac{k_B T z_0}{D_R m g} = \frac{9\eta z_0}{2\rho R^2 g} = \varphi \frac{z_0}{R^2} \quad , \qquad (1)$$

where the Stokes expression is used for the droplet diffusion constant $D_R = k_B T/(6\pi\eta R)$, the droplet mass is given by $m = 4\pi R^3 \rho/3$, and values for the viscosity of air $\eta$, water density $\rho$, thermal energy $k_B T$, and the gravitational constant $g$ are given in Table I. The numerical prefactor in Eq. (1) turns out to be $\phi = 0.85 \times 10^{-8}$ m s. For a droplet with radius $R$ = 5 μm placed initially at a height of $z_0$ = 2 m, the sedimentation time is $\tau_{sed}$ = 680 s = 11 minutes, other numbers are given in Table II. The droplet radius $R$ = 5 μm is often defined as a threshold radius below which the sedimentation time is sufficiently long to be considered relevant for infections. An exact calculation of the sedimentation time distribution is given in Appendix A, which shows that the relative standard deviation of the mean sedimentation time is small for droplet radii larger than $R$ = 10 nm. Thus, the mean sedimentation time, $\tau_{sed}$ in Eq. (1), is a good estimate of typical sedimentation times for all droplets with $R > 10$nm.

Inertial effects due to the acceleration of a droplet that is initially at rest occur over the momentum diffusion or acceleration time, which is

$$\tau_{acc} = \frac{mD_R}{k_B T} = \frac{2\rho R^2}{9\eta} = \xi R^2 \quad , \tag{2}$$

where the numerical prefactor is given by $\xi$ = 8.37 x 10$^6$ s m$^{-2}$. Even for large droplets with $R$=100 µm, the acceleration time is $\tau_{acc} = 0.1\ s$, showing that droplets rapidly reach their terminal velocity, so that acceleration effects can be neglected.

The lateral diffusion length during the time a droplet is sedimenting in stagnant air is readily estimated. For this, the mean-squared diffusion length at the mean sedimentation time is calculated from

$$x_{diff}^2 = 2D_R \tau_{sed} \ .$$

Inserting the mean sedimentation time from Eq. (1) results in

$$x_{diff} = \sqrt{\frac{k_B T z_0}{4R^3 \rho g}} \quad ,$$

which yields $x_{diff}$ = 0.3 mm for a droplet of radius $R$ = 1 µm and $x_{diff}$ = 1 cm for a droplet of radius $R$ = 100 nm. The lateral diffusion of a droplet during its sedimentation time is, therefore, very limited and will be dominated by the initial emission speed, air flow and turbulent convection effects.

## 3. Droplet evaporation without non-volatile solutes

So far, the effect of evaporation has been neglected, which decreases the droplet radius during its descent to the ground and therefore increases the sedimentation time. For evaporation of a droplet at rest, which defines the so-called stagnant-flow approximation, the time-dependent shrinking of the radius occurs in the diffusion-limited evaporation scenario, which is valid for radii larger than $R$= 100 nm, and is given by (see Appendices B and C)

$$R(t) = R_0(1 - \theta\ t(1 - RH)/R_0^2)^{1/2}\ . \tag{3}$$

Here $R_0$ is the initial droplet radius and the numerical prefactor is given by

$$\theta = 2D_w c_g v_w \left(1 - \frac{\varepsilon_C \varepsilon_T}{1+\varepsilon_C \varepsilon_T}\right) = 3.5 \times 10^{-10} m^2/s \quad , \tag{4}$$

where $\theta$ has units of a diffusion constant. The values for the water diffusion constant in air $D_w$, the liquid water molecular volume $v_w$ and the saturated water vapor concentration $c_g$ at room temperature 25 °C are given in Table I. $RH$ denotes the relative water humidity. The reduction of the water vapor concentration at the droplet surface due to evaporation cooling is described by the linear coefficient $\varepsilon_C$ according to $c_g^{surf} \approx c_g(1 - \varepsilon_C \Delta T)$. Here $c_g^{surf}$ denotes the water vapor concentration at the droplet surface, which has a temperature that is reduced compared to the ambient air (at temperature 25 °C) by $\Delta T$. The linear coefficient is given by $\varepsilon_C = 0.04$ (see Appendix C). The temperature reduction at the droplet surface is obtained by solving the coupled heat-flux and water diffusion-flux equations in a self-consistent manner and turns out to be linearly related to the relative humidity as $\Delta T = T_0 - T_s = \frac{\varepsilon_T(1-RH)}{1+\varepsilon_T \varepsilon_s} = 17(1 - RH)$, where the coefficient $\varepsilon_T$ is given by $\varepsilon_T \equiv \frac{D_w c_g h_{ev}}{\lambda_{air}} = 52$ (see Appendix C). Interestingly, at zero relative humidity ($RH = 0$), the droplet surface is cooled by 17 K, so while the cooling effect is quite significant, droplet freezing does not occur at room temperature. The factor in Eq. (4) that accounts for the evaporation cooling effect is given by $\left(1 - \frac{\varepsilon_C \varepsilon_T}{1+\varepsilon_C \varepsilon_T}\right) = 0.32$, so cooling considerably slows down the evaporation process and cannot be neglected (see Appendices B and C for the derivation of Eq. (3)). If the radius becomes smaller than 100 nm before the end of the drying process, a crossover to the reaction-rate limited evaporation regime takes place, as is discussed in Appendix D. For radii larger than 60 μm, the flow around the droplet speeds up the evaporation process and at the same time becomes non-Stokesian due to non-linear hydrodynamics effects, which can be treated analytically by double-boundary-layer theory including concentration and flow boundary layers, as discussed in Appendices E, F, G, H, I. Internal mixing effects inside the droplet are irrelevant for droplet radii below roughly 100 μm (see Appendix J). It transpires that the stagnant flow approximation used to derive Eq. (3) is valid for the initial radius range between 100 nm and 60 μm, which coincides with the range that produces the largest viral air load, as will be shown below.

From Eq. (3) it is seen that the decrease in the radius starts slowly and accelerates with time, it is therefore dominated by the initial stage of evaporation. Because of this, the time for evaporation down to a radius at which osmotic effects due to dissolved solutes and the presence of virion particles within the droplet balance the water vapor chemical potential, can be approximated as the time needed to reduce the droplet radius to zero, given by

$$\tau_{ev} = \frac{R_0^2}{\theta(1-RH)} \quad . \tag{5}$$

This relation has been first established based on empirical grounds by Wells (2). Notably, the evaporation time in Eq. (5) increases quadratically with the initial droplet radius $R_0$, while the sedimentation time in Eq. (1) decreases inversely and quadratically with the radius. Thus, at a relative humidity of $RH = 0.5$, a common value for room air, a droplet with an initial radius of $R_0 = 10$ μm has an evaporation time of $\tau_{ev} = 0.57$ s, but needs (neglecting the reduction of the radius) $\tau_a = 170$ s to sediment to the ground. Consequently, it will dry out and stay floating for an even longer time, depending on its final dry radius. Other numerical examples for evaporation times are given in Table II. A quick estimate of the critical radius below which droplets will completely dry out before sedimenting to the ground is obtained by equating the floating and evaporation times in Eqs. (1) and (5), which gives

$$R_0^{crit} \approx (\varphi \theta z_0 (1 - RH))^{1/4} \quad . \tag{6}$$

From an empirical analysis, this relation has also been established by Wells (2). For a relative humidity of *RH* = 0.5 and an initial height of $z_0$ =2m, the estimate $R_0^{crit} \approx 42 \ \mu m$ is obtained from Eq. (6). To accurately calculate the critical initial radius below which a droplet completely dries out before falling to the ground, one needs to take into account the decrease in droplet radius due to evaporation. Consequently, its diffusion constant and the gravitational force change during sedimentation. As detailed in Appendix B, the sedimentation time in the presence of a finite relative humidity *RH* < 1 is given by

$$\tau_{sed}^{RH} = \tau_{ev} \left[1 - \left(1 - \frac{2\varphi z_0}{\tau_{ev} R_0^2}\right)^{1/2}\right] \quad . \tag{7}$$

By inserting Eq. (5) into Eq. (7) it is seen that in the limit *RH* = 1 the result of Eq. (1) is recovered. The critical radius is defined by the initial radius for which the droplet radius just vanishes as it hits the ground, it follows from equating Eqs. (5) and (7) as

$$R_0^{crit} = (2\varphi \theta z_0 (1 - RH))^{1/4} \quad . \tag{8}$$

One obtains from Eq. (8) for *RH* = 0.5 and $z_0$ = 2 m the slightly higher estimate of $R_0^{crit} = 49 \ \mu m$ compared to the approximate result in Eq. (6). All droplets smaller than $R_0^{crit} = 49 \ \mu m$ will dry out before they hit the ground. In the absence of non-volatile solutes, the droplets will thus disappear for radii smaller than $R_0^{crit}$; in the presence of non-volatile solutes, the droplets can only shrink down to a radius that is predominantly determined by the solute content, as will be discussed in Section 4. In airliners the relative humidity is substantially lower than 0.5; in fact, for completely dry air with *RH* = 0, the critical radius predicted by Eq. (8) increases to $R_0^{crit} = 59 \ \mu m$. Note that the results presented here hold in still air; in air-conditioned rooms, convection due to air circulation will prevent some droplets from falling to the ground for a long time. Figure 1 shows droplet sedimentation times $\tau_{sed}^{RH}$ as a function of the initial radius $R_0$ according to Eq. (7) for an initial height of $z_0$ = 2 m for different relative humidities. In the limit *RH* = 1 no evaporation takes place and the result of Eq. (1) is recovered (thick black line). As the initial radius approaches the critical radius $R_0^{crit}$, given by Eq. (8) and indicated by a broken line, the droplet disappears. The thin solid colored lines denote the evaporation times according to Eq. (5), the crossing of the evaporation and sedimentation times happens at the critical radius. The qualitative shape of these curves has been empirically established by Wells (2).

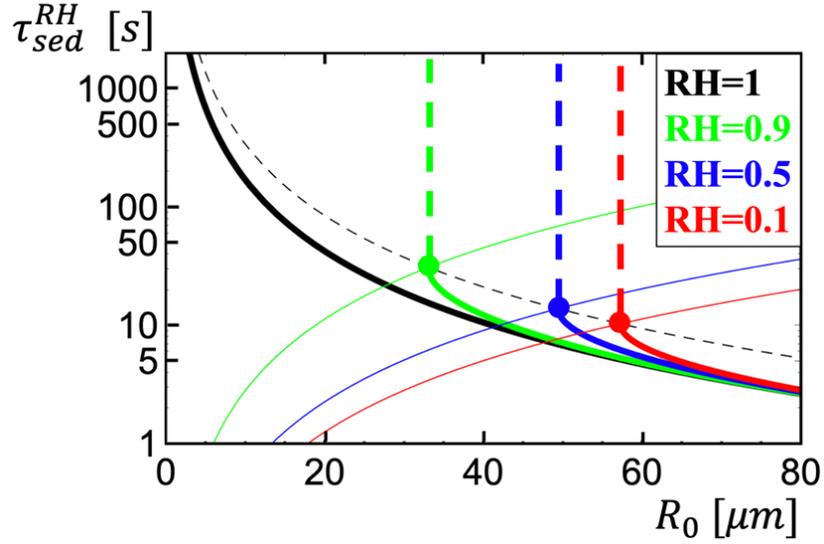

**Fig. 1:** Sedimentation time of droplets $\tau_{sed}^{RH}$ in the presence of evaporation as a function of the initial radius $R_0$ in the absence of non-volatile solutes according to Eq. (7) for an initial height of $z_0$ = 2 m. Results are shown for different relative humidities, in the limit RH = 1 no evaporation takes place and the result in Eq. (1) is recovered (thick black line). As the initial radius approaches the critical radius $R_0^{crit}$, given by Eq. (8) and indicated by a black broken line, the droplet disappears (indicated by vertical broken lines). The thin solid colored lines denote the evaporation time Eq. (5).

## 4. Droplet evaporation in the presence of non-volatile solutes

So far, the presence of non-volatile solutes in the initial droplet, which produces a lower limit for the droplet radius that can be reached by evaporation, has been neglected. Saliva contains a volume percentage of about 99.5 % water (29), the radius of a saliva droplet thus can maximally shrink by a factor $200^{1/3}$ = 5.8. Some of the water will stay inside the final droplet because of hydration effects. Assuming that the final state keeps 50% strongly bound hydration water, the droplet can thus maximally shrink by a factor of $100^{1/3}$ = 4.6. It is important to note that the concentration of non-volatile solutes (including virions) has in this explicit numerical example increased by a factor of 100 due to evaporation. Solutes in the droplet decrease the water vapor pressure, and therefore limit the equilibrium droplet radius that is obtained in the long-time limit according to (see Appendix K)

$$R_{ev} = R_0 \left(\frac{\Phi_0}{1-RH}\right)^{1/3} . \qquad (9)$$

Here, $R_0$ is the initial radius and $\Phi_0$ is the initial volume fraction of solutes, including strongly bound hydration water. Only for RH = 0 does a droplet dry out to the minimal possible radius of $R_{ev} = R_0(\Phi_0)^{1/3}$; for finite relative humidity the equilibrium droplet radius is characterized by an equilibrium solute volume fraction of $\Phi_{ev} = 1 - RH$. As an example, for RH = 0.5, the free water and solute (including hydration water) volume fractions in the equilibrium state equal each other. Equation (9) is modified for solutes that perturb the water activity, but for most solutes non-ideal water solution effects can be neglected.

Taking into account the water vapor-pressure reduction during the evaporation process, the analytical result for the radius-dependent evaporation time, which is the time it takes for the droplet radius to decrease from its initial value $R_0$ to $R$, is given by

$$t(R) = \frac{R_{ev}^2}{\theta(1-RH)}\left[\mathcal{L}\left(\frac{R_0}{R_{ev}}\right) - \mathcal{L}\left(\frac{R}{R_{ev}}\right)\right] \quad , \tag{10}$$

as derived in Appendix K. A very accurate yet simple approximation for the scaling function $\mathcal{L}(x)$ is

$$\mathcal{L}\left(\frac{R}{R_{ev}}\right) \cong \left(\frac{R}{R_{ev}}\right)^2 + \frac{2}{3}\ln\left(1 - \frac{R_{ev}}{R}\right) \quad , \tag{11}$$

so Eq. (10) can be written as

$$t(R)/\tau_{ev} = 1 - \frac{R^2}{R_0^2} - \frac{2R_{ev}^2}{3R_0^2}\ln\left(\frac{R_0(R-R_{ev})}{R(R_0-R_{ev})}\right) \quad , \tag{12}$$

where $\tau_{ev}$ denotes the evaporation time to shrink the solute-free droplet down to a vanishing radius from Eq. (5). This expression clearly demonstrates the logarithmic osmotic slowing down of the evaporation process due to the decreasing droplet water concentration as the droplet radius $R$ approaches the equilibrium droplet radius $R_{ev}$. Neglecting this kinetic slowing down, which is represented by the last term in Eq. (12), one obtains the limiting result

$$t(R)/\tau_{ev} = 1 - \frac{R^2}{R_0^2} \quad , \tag{13}$$

from which an approximate expression for the evaporation time in the presence of solutes follows as

$$\tau_{ev}^{sol} = \tau_{ev}\left(1 - \frac{R_{ev}^2}{R_0^2}\right) \quad , \tag{14}$$

which for small initial solute concentrations represents a rather small correction to the evaporation time given by Eq. (5).

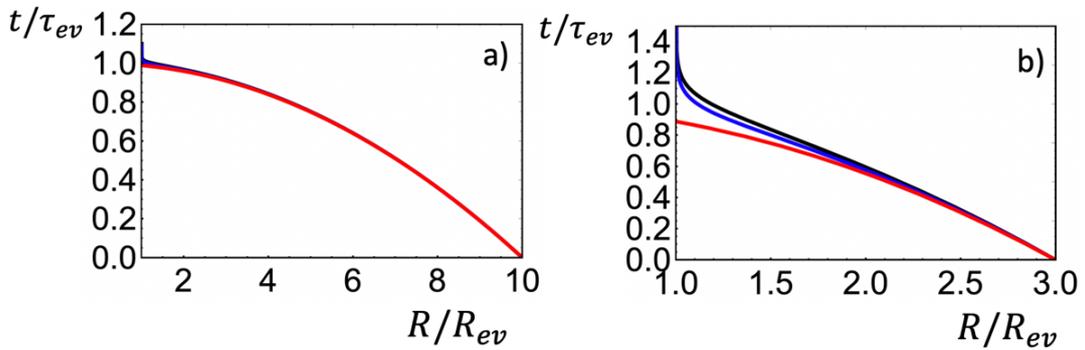

**Fig. 2:** Scaling plot of the evaporation time $t(R)$ as a function of the droplet radius $R$ in the presence of solutes according to Eq. (10) (black lines). In a) the ratio of the initial droplet radius to the final equilibrium radius is $\frac{R_0}{R_{ev}} = 10$ and in b) this ratio is $\frac{R_0}{R_{ev}} = 3$. The red lines show the evaporation time when the water vapor pressure

reduction is neglected, Eq. (13); the blue lines show the approximation Eq. (12). The solute-induced water-vapor pressure reduction becomes significant only for radii close to the final equilibrium radius $R_{ev}$ and leads to a diverging evaporation time.

Figure 2 shows the rescaled evaporation time as a function of the reduced droplet radius according to Eq. (10) as black lines. The presence of solutes only becomes relevant for droplet radii that are close to the final equilibrium radius $R_{ev}$ and gives rise to a divergent evaporation time. Except for this final stage of evaporation, the formula Eq. (13) (red lines) describes the evaporation very accurately and will be used for all further calculations.

The sedimentation of not too large droplets thus can approximately be split into two stages: In the first stage, the droplets shrink down to a radius given by Eq. (9), and in a second stage the droplets sediment for an extended time with a fixed radius. The total sedimentation time follows as (see Appendix K)

$$\tau_{sed}^{sol} = \frac{\varphi z_0}{R_{ev}^2} - \frac{\tau_{ev}}{2}\left(\frac{R_0}{R_{ev}} - \frac{R_{ev}}{R_0}\right)^2. \quad (15)$$

For droplets that are so large that they do not reach the radius $R_{ev}$ before they hit the ground, Eq. (7) describes the sedimentation time very accurately.

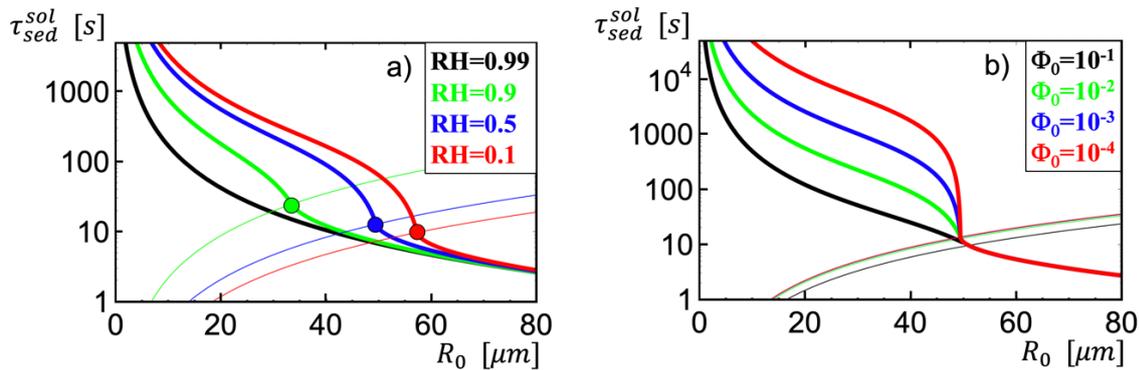

**Fig. 3:** a) Sedimentation time of droplets $\tau_{sed}^{sol}$ as a function of the initial radius $R_0$ in the presence of non-volatile solutes with initial volume fraction $\Phi_0 = 0.01$ (which includes strongly bound hydration water) according to Eq. (15), for an initial height of $z_0$ = 2 m. Results are shown for different relative humidities, in the case *RH* = 0.99 no evaporation takes place and the result Eq. (1) is recovered (thick black line). The thin solid colored lines denote the evaporation time Eq. (14). For small droplet radii sedimentation is a two-stage process and droplets first evaporate down to the equilibrium radius $R_{ev}$ and then stay floating in air for an extended time. b) Sedimentation time of droplets $\tau_{sed}^{sol}$ as a function of the initial radius $R_0$ for fixed relative humidity *RH* = 0.5 and an initial height of $z_0$ = 2 m in the presence of non-volatile solutes with different initial volume fractions $\Phi_0$ according to Eq. (15).

In Fig. 3a the droplet sedimentation time $\tau_{sed}^{sol}$ is plotted as a function of the initial radius $R_0$ in the presence of non-volatile solutes with an initial solute volume fraction $\Phi_0 = 0.01$ and an initial height of $z_0$ = 2 m according to Eq. (15) for a few different relative humidities. For *RH* = 0.99 no evaporation takes place and the result of Eq. (1) is recovered (thick black line). The thin solid colored lines denote the evaporation time Eq. (14). For small droplet radii sedimentation is a two-stage process; droplets first evaporate down to the equilibrium radius $R_{ev}$ and then stay floating in air for an extended time. Large droplets do not reach $R_{ev}$ before they hit the ground, the transition between these two scenarios is illustrated by filled circles. In Fig. 3b the droplet sedimentation time $\tau_{sed}^{sol}$ is plotted for fixed relative humidity *RH* = 0.5 and different initial solute volume fractions $\Phi_0$. Figure 3 illustrates that the sedimentation times are significantly increased due to evaporation. In fact, as shown in Table II, for a relative humidity $RH = 0.5$ and $\Phi_0 = 0.01$, the sedimentation times of droplets increase for not too large radii by more than a factor of 10 due to evaporation.

**5. Steady-state number of virions sedimenting in air**

The virion content of a droplet is proportional to its initial volume. Denoting the droplet production rate of a single human who is speaking, which in principle depends on droplet radius, as $f_{drop}$, the number of humans that are simultaneously speaking as $m$, the virion number concentration in saliva as $c_{vir}$, the total number of virions sedimenting in air denoted as $N_{vir}$, is in the steady state given by

$$N_{vir} = \frac{4\pi R_0^3}{3} \tau_{sed}^{sol} \, m f_{drop} \, c_{vir} \, . \qquad (16)$$

In Figure 4a, the product of the initial droplet volume $\frac{4\pi R_0^3}{3}$ and the sedimentation time $\tau_{sed}^{sol}$, which appears in Eq. (16) on the right side, is plotted as a function of the initial droplet radius for a few different relative humidities. This quantity is for $RH = 0.5$ broadly peaked and rather constant for initial droplet radii between 10 μm and 40 μm. This interesting property is due to the fact that smaller droplets contain less volume but evaporate faster and thus have a longer sedimentation time. This means that the precise dependence of the droplet production rate $f_{drop}$ on the initial droplet radius $R_0$ is not very important; the only important quantity is the total rate of droplets produced in the radius range between 10 μm and 40 μm.

The concentration of SARS-CoV-2 viruses in saliva can be assumed to be $c_{vir} = 10^6 \, ml^{-1}$, which is a conservative estimate given the recent measurement of viral RNA concentration in human sputum, which yielded a value of $7 \times 10^6 \, ml^{-1}$ (28). The droplet production rate from speaking was recently estimated in the droplet radius range between 12 μm and 21 μm as $2.6 \times 10^3 \, s^{-1}$ (23) and in the radius range higher than about 20 μm as $\sim 10^3 \, s^{-1}$ (22), from which the conservative estimate $f_{drop} \approx 10^3 \, s^{-1}$ is constructed. Together this gives a factor $f_{drop} c_{vir} = 10^9 \, s^{-1} ml^{-1} = 10^{-3} \, s^{-1} \mu m^{-3}$. For a single human (*m=1*), this factor results in a steady-state number of virions floating in air between $10^4$ and $10^5$ for a humidity value around of *RH = 0.5*, as seen in Figure 4a on the right scale. This estimate assumes that the person does not wear a mask and is constantly speaking, obviously, it will be reduced if the person speaks only intermittently.

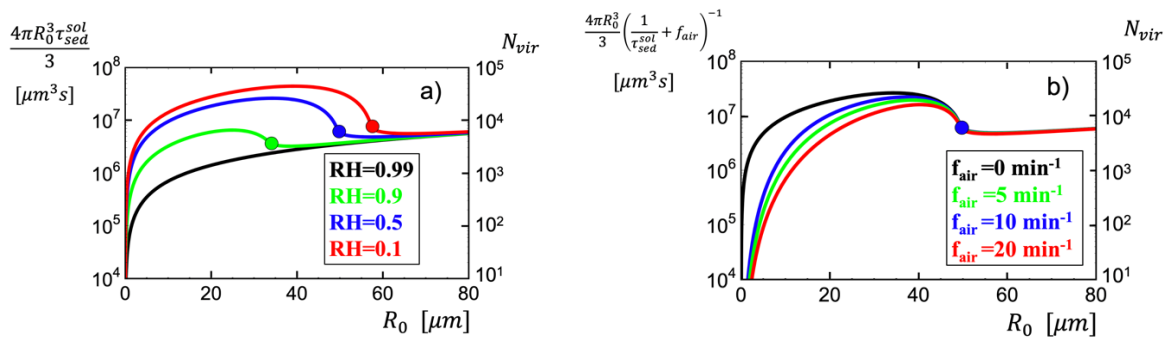

**Fig. 4:** a) Product of the sedimentation time of droplets $\tau_{sed}^{sol}$ and the initial droplet volume $\frac{4\pi R_0^3}{3}$ as a function of the initial radius $R_0$ in the presence of non-volatile solutes as given by Eq. (15) for an initial height of $z_0$ = 2 m and an initial solute volume fraction $\Phi_0 = 0.01$. Results are shown for different relative humidities, in the case $RH = 0.99$ no evaporation takes place and the result Eq. (1) is recovered (thick black line). The right scale shows the steady-state number of virions $N_{vir}$ sedimenting in air assuming droplet production at a rate $f_{drop} = 10^3 \, s^{-1}$ for a single droplet producer (m = 1) and for a saliva virion concentration $c_{vir} = 10^6 \, ml^{-1}$ according to Eq. (16). b) Same as a) but including the effect of air exchange with a rate $f_{air}$ according to Eq. (19). Results are shown for RH=0.5 and for four different air change rates in a closed room, assuming well-mixed air and a single droplet producer m=1.

In open air, the produced droplets will dilute due to the producing person moving around and due to wind and convection effects. The open-air scenario is typically considered harmless. The situation in closed rooms is very different. The differential equation that describes the time-dependent number of droplets in a room is given by

$$\frac{dN_{drop}(t)}{dt} = m f_{drop} - \frac{N_{drop}(t)}{\tau_{sed}^{sol}} - N_{drop}(t) f_{air} \quad . \quad (17)$$

The first term on the right side is the droplet production term, proportional to the droplet production rate $f_{drop}$ and the number of droplet producers $m$; the second term is the droplet loss rate due to sedimentation to the ground; and the last term is the droplet loss rate due to air exchange that is proportional to the air-exchange rate $f_{air}$. In writing the last term, the assumption is made that the room air is well mixed, which should be a good approximation in ventilated rooms and for sedimentation times that exceed a minute (valid for initial droplet radii above 40 µm). Recommended air-exchange rates range from $f_{air} = 5/h$ in residential rooms up to $f_{air} = 20/h$ in multiply occupied offices and restaurants. In a steady state, the droplet number does not change with time and from Eq. (17) follows as

$$N_{drop} = m f_{drop} \left(\frac{1}{\tau_{sed}^{sol}} + f_{air}\right)^{-1} \quad . \quad (18)$$

Thus, the total number of virions sedimenting in air follows as

$$N_{vir} = \frac{4\pi R_0^3}{3} c_{vir} N_{drop} = \frac{4\pi R_0^3}{3} c_{vir} m f_{drop} \left(\frac{1}{\tau_{sed}^{sol}} + f_{air}\right)^{-1}, \quad (19)$$

which is a generalization of Eq. (16). The effect of a finite air-exchange rate reduces the total number of virions floating in air significantly but not completely, as is seen in Fig. 4b. In particular, the virion number from droplets with radii between $R_0 = 20 \, \mu m$ and $R_0 = 40 \, \mu m$ is not affected much by a finite air-exchange rate, this is so because the inverse sedimentation time in this range is of the order of the air-exchange rate and thus mitigates the air-exchange efficiency. Air recirculation between different rooms without air exchange is a further risk, as it distributes the virion air load between all ventilated rooms.

An important question for infection risk estimates is the number of virions that are inhaled by a person per minute. Denoting the tidal volume in normal breathing as $V_{tidal}$, the average respiratory frequency as $f_{resp}$, the volume of a closed room as $V_{room}$, the rate at which virions are inhaled by a person is given by

$$f_{inhale} = \frac{f_{resp} V_{tidal}}{V_{room}} N_{vir} = \frac{f_{resp} V_{tidal}}{V_{room}} \frac{4\pi R_0^3}{3} c_{vir} \, m \, f_{drop} \left(\frac{1}{\tau_{sed}^{sol}} + f_{air}\right)^{-1}, \quad (20)$$

where again the well-mixing assumption for air is used. The tidal breathing volume of adults is about $V_{tidal} = 0.5 \, l$ and the average respiratory frequency is about $f_{resp} = 20/min$. Assuming a room volume corresponding to an area of 20 square meters and a height of 2 m, resulting in $V_{room} = 4 \times 10^4 \, l$, the prefactor in Eq. (20) comes out as $\frac{f_{resp} V_{tidal}}{V_{room}} = 2.5 \times 10^{-4} \, min^{-1}$. As seen in Fig. 4, the steady-state number of sedimenting virions is even for a single speaker (m = 1) larger than $N_{vir} \approx 10^4$ in the entire radius range between $R_0 = 10 \, \mu m$ and $R_0 = 40 \, \mu m$ for a typical relative humidity RH = 0.5, and is only weakly reduced by increased air-exchange rates, as demonstrated in Fig. 4b. The conclusion is that droplets produced by a constantly-speaking single person give rise to a virion inhalation rate of a passive bystander of at least $f_{inhale} = 2.5 \, min^{-1}$ in a wide droplet radius range.

It is, of course, not straightforward to derive the infection risk from the virion inhalation frequency. It is known that SARS-CoV-2 viruses remain viable in aerosols for at least 3 hours (24), which is longer than the sedimentation times in the relevant radius range, as seen in Fig. 3. As a comparison, on an inanimate surface these viruses stay infectious for days (24; 30). As a further complication, the relative humidity seems to have a significant influence on virus stability, it was shown for bacteriophages and influenza viruses that stability is minimal at intermediate humidities around RH = 0.5 and is increased both for lower and larger humidities (31; 32). Unfortunately, similar data is not yet available for SARS-CoV-2 viruses. Many factors determine the likelihood that a virus will spread from one person to another and that disease will result, but for other viruses it is known that inhaling as few as 5 virions can cause infection (33), so the above estimate of a virion inhalation rate of $f_{inhale} = 2.5 \, min^{-1}$, which is a conservative estimate, should be relevant for the assessment of the viral airborne infection risk.

## 6. Discussion and conclusion

From the above analysis, it is clear that droplet sedimentation is a complex problem. In order to come up with analytical predictions a number of simplifying assumptions had to be made. It has been assumed that diffusion within the droplet happens quickly enough, so that the water concentration at the droplet surface does not differ significantly from the mean water concentration in the droplet. In Appendix J it is shown that this approximation is valid for radii below R = 100 μm, which coincides with the relevant radius range for airborne infections. Surface tension effects, which increase the water vapor pressure, are negligible for droplets with radii larger than R = 1 nm, as explained in Appendix L. Likewise, the pressure increase due to evaporation and the change of droplet mass density with evaporation has been neglected.

Human sneeze was shown to produce a turbulent gas cloud of droplets mixed with hot and moist exhaled air, which can travel up to 8 m (34). It was demonstrated that the warm atmosphere in this cloud slows down evaporation for droplets that are small enough to reside inside the cloud for an extended time (35). The results presented here in principle hold also for droplets produced by sneezing once the droplets have left the sneeze cloud.

Droplets larger than $R$ = 100 μm quickly fall to the ground, but can spread disease by ballistically landing on other people or on surfaces, which is a distinct infection mechanism and not considered here.

In summary, the evaporation of aqueous droplets with initial radii 100 nm < $R_0$ < 60 μm, which includes the radius range relevant for air-borne infection pathways, can be described by the stagnant air approximation in the diffusion limit. These calculations demonstrate in terms of analytical formulas that droplets in the entire range of radii below $R_0^{crit} = 49\ \mu m$ for $RH$ = 0.5, shrink significantly from evaporation before they fall to the ground and thus stay floating in air longer than their initial radius would suggest. This leads to an enhanced viral air load for droplets in the entire initial radius range 10 μm < $R_0$ < 40 μm, which is exactly the radius range of droplets primarily produced by speaking (22) (23). A simple estimate of the viral inhalation frequency in a closed room suggests that 2.5 virions are inhaled per minute if one person is constantly speaking and not wearing a mask, typical air-exchange rates do not lower this number significantly. Thus speaking and presumably more so singing are shown to increase the risk of airborne viral infections substantially, which can be reduced efficiently by wearing a mouth cover (22) (23). The analytical formulas presented in this work will in the future facilitate further calculations of droplet dwell times that include convection and turbulent flow effects.

**Table I:** List of numerical constants used.

| Symbol | Description | Value |
| --- | --- | --- |
| $k_B T$ | thermal energy | 4.1 x 10$^{-21}$ J at 25°C |
| $\eta$ | viscosity of air | 1.86 x 10$^{-5}$ kg/ms at 25°C |
| $\eta$ | viscosity of air | 1.73 x 10$^{-5}$ kg/ms at 0°C |
| $\rho$ | liquid water density | 10$^3$ kg/m$^3$ |
| $g$ | gravitational constant | 9.81 m/s$^2$ |
| $D_w$ | water diffusion constant in air | 2.82 x 10$^{-5}$ m$^2$/s at 25°C |
| $D_w$ | water diffusion constant in air | 2.2 x 10$^{-5}$ m$^2$/s at 0°C |
| $D_w^l$ | water diffusion constant in liquid water | $2.3 \times 10^{-9} m^2/s$ at 25°C |
| $m_w$ | water molecular mass | 2.99 x 10$^{-26}$ kg |
| $v_w$ | liquid water molecular volume | 3.07 x 10$^{-29}$ m$^3$ at 25°C |
| $v_w$ | liquid water molecular volume | 2.99 x 10$^{-29}$ m$^3$ at 4°C |
| $c_g$ | saturated vapor water concentration | 6.6 x 10$^{23}$ m$^{-3}$ at 25°C |
| $P_{vap}$ | water vapor pressure | 2340 Pa at 20°C |
| $P_{vap}$ | water vapor pressure | 610 Pa at 0°C |
| $\rho_{air}$ | density of air | 1.18 x kg m$^{-33}$ at 25°C |
| $\nu$ | kinematic air viscosity | 1.5 x 10$^{-5}$ m$^2$/s at 25°C |
| $k_c$ | condensation reaction rate coefficient | 300 m/s |
| $a_{air}$ | air temperature diffusivity | 2 x 10$^{-5}$ m$^2$/s |
| $a_w$ | liquid water temperature diffusivity | 1.4 x 10$^{-7}$ m$^2$/s |
| $h_{ev}$ | molecular evaporation enthalpy of water | $7.3 \times 10^{-20}\ J$ at 25°C |
| $h_{ev}$ | molecular evaporation enthalpy of water | $7.1 \times 10^{-20}\ J$ at 0°C |
| $h_m$ | molecular melting enthalpy of water | $1.0 \times 10^{-20}\ J$ |
| $C_P^l$ | molecular heat capacity of liquid water | $1.3 \times 10^{-22}\ J$ at 25°C |
| $\lambda_{air}$ | heat conductivity of air | 0.026 W/mK at 25°C |
| $\lambda_{air}$ | heat conductivity of air | 0.024 W/mK at 0°C |

**Table II:** List of representative sedimentation and evaporation times. $R_0$ denotes the initial droplet radius. $\tau_{sed}$ (*RH*=1) is the sedimentation time from a height of 2 meters without evaporation. $\tau_{ev}$ (*RH*=0.5) is the evaporation time at a relative humidity of *RH*=0.5 in the absence of non-volatile solutes in the droplet. $\tau_{sed}^{RH}$ (*RH*=0.5) is the sedimentation time in the absence of non-volatile solutes at a relative humidity of RH=0.5 from a height of 2 meters. $\tau_{sed}^{sol}$ (*RH*=0.5) is the sedimentation time from a height of 2 meters at a relative humidity of *RH*=0.5 in the presence of an initial volume fraction $\Phi_0 = 0.01$ of non-volatile solutes in the droplet.

| $R_0\ [\mu m]$ | 1 | 2.5 | 5 | 10 | 20 | 30 | 40 | 50 |
|---|---|---|---|---|---|---|---|---|
| $\tau_{sed}\ (RH=1)$ | 5 h | 45 min | 11 min | 170 s | 43 s | 19 s | 11 s | 7 s |
| $\tau_{ev}\ (RH=0.5)$ | 0.0057 s | 0.036 s | 0.14 s | 0.57 s | 2.28 s | 5.14 s | 9.14 s | 14.3 s |
| $\tau_{sed}^{RH}\ (RH=0.5)$ | $\infty$ | $\infty$ | $\infty$ | $\infty$ | $\infty$ | $\infty$ | $\infty$ | 11.2 s |
| $\tau_{sed}^{sol}\ (RH=0.5)$ | 64 h | 10 h | 154 min | 38 min | 9 min | 226 s | 91.0 s | 11.2 s |

**Acknowledgements:** I thank William Eaton for helpful comments and for bringing this problem to my attention. Discussions with A. Bax, L. Bocquet, J.-F. Dufreche, I. Gersonde, P. Loche, D. Lohse are gratefully acknowledged. This research has been funded by Deutsche Forschungsgemeinschaft (DFG) through grant CRC 1114 "Scaling Cascades in Complex Systems", grant Number 235221301, Project C02 and by the ERC Advanced Grant No. 835117.

**Appendix A: Diffusive droplet sedimentation without evaporation**

The density distribution of droplets that diffuse in a viscous medium (such as air) under the influence of gravitational force is given by the diffusion equation

$$\frac{d}{dt} P(z,t) = D_R \frac{d^2}{dz^2} P(z,t) + V \frac{d}{dz} P(z,t) \,, \tag{A1a}$$

where the stationary velocity is defined as

$$V = \frac{D_R m g}{k_B T} \tag{A1b}$$

and $D_R$ is the droplet diffusion coefficient, $m$ the droplet mass, $g$ the gravitational acceleration, $k_B T$ the thermal energy and $P(z,t)$ is the density of droplets at height $z$ and time $t$. The fact that droplets do not return to air once they reach the ground at height $z=0$ is accounted for by a vanishing density distribution at the ground, $P(z=0, t) = 0$, which is the absorbing boundary condition. The Laplace-transformed density distribution at time $t$, given that at time $t = 0$ droplets are placed at height $z_0$, the so-called Green´s function, is given by

$$\tilde{P}(\omega, z|z_0) = \frac{(e^{\gamma_1 z_0} - e^{-\gamma_2 z_0})\, e^{-\gamma_1 z}}{\sqrt{V^2 + 4 D_R \omega}} \quad \text{for } z > z_0 \tag{A2a}$$

$$\tilde{P}(\omega, z|z_0) = \frac{e^{-\gamma_2 z_0}(e^{\gamma_2 z} - e^{-\gamma_1 z})}{\sqrt{V^2 + 4 D_R \omega}} \quad \text{for } z < z_0 \quad , \tag{A2b}$$

where the decay lengths are defined as

$$\gamma_1 = \sqrt{\frac{V^2}{4 D_R^2} + \frac{\omega}{D_R}} + \frac{V}{2 D_R}$$

$$\gamma_2 = \sqrt{\frac{V^2}{4D_R^2} + \frac{\omega}{D_R}} - \frac{V}{2D_R}.$$

The survival fraction of droplets, i.e. the fraction of droplets that have not yet reached the ground, is obtained by the integral over the entire density distribution and given by

$$\tilde{S}(\omega, z_0) = \int_0^\infty dz\, \tilde{P}(\omega, z|z_0) = \left(\frac{4D_R}{V^2}\right) \frac{1 - e^{-\tilde{z}_0[\sqrt{1+\tilde{\omega}}-1]/2}}{\tilde{\omega}}, \quad (A3)$$

where rescaled variables $\tilde{z}_0 = \frac{z_0 V}{D_R}$ and $\tilde{\omega} = \frac{4\omega D_R}{V^2}$ are introduced. The inverse Laplace transform reads in closed form

$$S(t, z_0) = \frac{2D_R}{V^2}\left[1 - erf\left(\tilde{t}^{\frac{1}{2}} - \frac{\tilde{z}_0 \tilde{t}^{-\frac{1}{2}}}{4}\right) - e^{\tilde{z}_0} erfc\left(\tilde{t}^{\frac{1}{2}} - \frac{\tilde{z}_0 \tilde{t}^{-\frac{1}{2}}}{4}\right)\right], \quad (A4a)$$

where $\tilde{\omega}\tilde{t} = \omega t$, and has for large times the asymptotic decay

$$S(t, z_0) \sim z_0 t^{-3/2} e^{-V^2 t/4D_R}. \quad (A4b)$$

This shows that all higher moments exist. The absorption or sedimentation time distribution is given by

$$K(t, z_0) = -\frac{\partial S(t, z_0)}{\partial t} \quad (A5)$$

and is normalized. The first moment of the absorption distribution, the mean absorption or sedimentation time, is given by

$$\tau_{sed} = \int_0^\infty dt\, t K(t, z_0) = \int_0^\infty dt\, S(t, z_0) = \tilde{S}(\omega, z_0)|_{\omega \to 0}. \quad (A6)$$

Likewise, the second moment is given by

$$\tau_a^{(2)} = \int_0^\infty dt\, t^2 K(t, z_0) = 2\int_0^\infty dt\, t S(t, z_0) = -2\frac{\partial \tilde{S}(\omega, z_0)}{\partial \omega}\Big|_{\omega \to 0}. \quad (A7)$$

The explicit result for the mean adsorption or sedimentation time therefore reads

$$\tau_{sed} = \frac{z_0}{V} = \frac{k_B T z_0}{D_R m g} \quad (A8)$$

and is the average time for a droplet falling with constant velocity *V* to reach the ground.

The thermal equilibrium mean-height of a droplet above the ground (in the absence of absorption) is from the equipartition theorem given by

$$z_{eq} = \frac{k_B T}{mg}. \quad (A9)$$

Using $m = 4\pi R^3 \rho/3$ and numerical constants from Table I, for droplet radii $R$ = 1 nm, 10 nm, 100 nm equilibrium heights of $z_{eq} = 100$ m, 100 mm, 100 μm are obtained, so it is seen that thermal effects can be safely neglected for all but the smallest droplets. The relative standard deviation of the absorption or sedimentation time follows from Eq. (A7) as

$$\frac{\Delta \tau_a}{\tau_a} = \frac{\sqrt{\tau_a^{(2)} - \tau_a^2}}{\tau_a} = \sqrt{\frac{2z_{eq}}{z_0}} \quad . \tag{A10}$$

Together with the result Eq. (A9), the relative standard deviation is seen to be small for droplets larger than $R$ = 10 nm and for an initial height $z_0$ in the meter range.

**Appendix B: Stagnant droplet evaporation in the diffusion-limited regime without evaporation cooling effects**

In this appendix, convection effects in the air around the droplet due to the finite speed of a falling droplet will be neglected, which increase the speed of evaporation and will be treated in Appendices E-H. Also, evaporation cooling effects will be neglected. The water vapor concentration around a spherical droplet at rest is described by the diffusion equation in radial coordinates

$$\frac{d}{dt}c(r,t) = \frac{D_w}{r^2}\frac{d}{dr}r^2\frac{d}{dr}c(r,t) \;, \tag{B1}$$

where $D_w$ denotes the molecular water diffusion constant in air. The stationary density distribution is given by

$$c(r) = c_0\left(1 + \frac{b}{r}\right) \;, \tag{B2}$$

where $c_0$ is the ambient water vapor concentration. Here the adiabatic approximation is used and the time it takes for the stationary distribution to build up, which can be shown to be small, is neglected. Particle conservation together with the reactive boundary condition at the droplet surface $r = R$ gives for the flux density $j$

$$j = -D_w\frac{d}{dR}c(R) = k_e c_l - k_c c(R) \;, \tag{B3}$$

where $k_e$ and $k_c$ are the evaporation and condensation reaction rate coefficients, which have units of velocity, and $c_l$ is the liquid water concentration inside the droplet (or, to be more precise, at the droplet surface). Inserting the solution Eq. (B2) into Eq. (B3), the resulting equation can be solved for the coefficient $b$ and the total water evaporation flux, $J$, is obtained as

$$J = 4\pi R^2 j = 4\pi D_w c_0 b = 4\pi R^2 D_w \frac{k_e c_l - k_c c_0}{D_w + k_c R} \quad . \tag{B4}$$

For saturated water vapor with concentration $c_g$, the evaporation flux must vanish, i.e. $k_e c_l = k_c c_g$, and the evaporation rate coefficient $k_e$ can be eliminated from Eq. (B4) to give

$$J = 4\pi R^2 D_w \frac{k_c c_g (1-RH)}{D_w + k_c R} \quad , \tag{B5}$$

where the relative fractional air humidity is defined as $RH = c_0/c_g$. As expected, the evaporation flux vanishes for $RH = 1$ corresponding to water-saturated air. The condensation reaction rate coefficient $k_c$ is large, since every water molecule that hits the air-water interface basically sticks. From molecular kinetic considerations, it follows that $k_c$ is given by the thermal molecular water velocity

$$k_c = \sqrt{\frac{k_B T}{m_w}} \approx 300 \; m/s \;, \tag{B6}$$

where the water molecule mass $m_w$ from Table I was used. The diffusion-limited rate scenario is defined by $k_c R > D_w$, which is realized for droplet radii $R > \frac{D_w}{k_c} = 100 \; nm$, where the water molecular diffusion constant in air, $D_w$ at 25°C from Table I, was used. In this limit, one can neglect the term proportional to $D_w$ in the denominator of Eq. (B5) and obtain the classical diffusion-limited result for the evaporation flux, which is linearly proportional to the droplet radius,

$$J = 4\pi R D_w c_g (1 - RH) \quad . \tag{B7}$$

Mass conservation of the droplet means that the evaporation flux is balanced by a decreasing radius, which can be written in terms of the droplet volume as

$$\frac{d}{dt}\left(\frac{4\pi}{3} R^3(t)\right) = -v_w J = -4\pi R D_w c_g v_w (1 - RH) \quad , \tag{B8}$$

where $v_w$ is the volume of a water molecule in the liquid phase. The differential equation (B8) is easily solved with the result

$$R(t) = R_0 \left(1 - t \frac{2 D_w c_g v_w (1-RH)}{R_0^2}\right)^{1/2} = R_0 (1 - \theta \, t (1 - RH)/R_0^2)^{1/2} \;, \tag{B9}$$

where $R_0$ is the initial droplet radius and the numerical prefactor is given by

$$\theta = 2 D_w c_g v_w = 1.1 \times 10^{-9} m^2/s \tag{B10}$$

and has units of a diffusion constant. Note that the calculation neglects cooling effects from evaporation, which substantially change the numerical prefactor, as shown in Appendix C. The water molecular volume in the liquid phase, $v_w$, and the water concentration of saturated water vapor, $c_g$, have been taken from Table I. It is seen that the shrinking of the radius starts slowly and accelerates over time. The evaporation time down to a radius where osmotic effects due to dissolved solutes and the presence of virions inside the droplet balance the evaporation chemical potential, can thus be approximated as the time needed to shrink the droplet radius to zero, given by

$$\tau_{ev} = \frac{R_0^2}{\theta (1-RH)} \quad . \tag{B11}$$

Notably, the evaporation time in Eq. (B11) increases quadratically with the initial droplet radius $R_0$, while the absorption time in Eq. (1) decreases inversely and quadratically with $R_0$. Thus, at a relative humidity of $RH = 0.5$, a common value for room air, a droplet with an initial radius of $R_0 = 1$ μm has an evaporation time of $\tau_{ev} = 1.8$ ms, but takes (neglecting shrinkage of the radius) $\tau_a = 5$ h to fall to the ground, so it will dry out and basically stay floating for an even longer time, depending on its final dry radius.

To calculate the critical initial radius below which a droplet completely dries out before falling to the ground, Eq. (1) is rewritten in terms of the instantaneous, radius-dependent droplet velocity $v(t)$ and combined with Eq. (B9) gives

$$v(t) = \frac{D_R g m(t)}{k_B T} = \frac{R^2(t)}{\varphi} = \frac{R_0^2}{\varphi}(1 - \theta t(1-RH)R_0^{-2}) = \frac{R_0^2}{\varphi}(1 - t/\tau_{ev}). \quad (B12)$$

The distance by which the droplet falls during time $t$ follows by integration as

$$\Delta z = \int_0^t dt'\, v(t') = \frac{R_0^2 \tau_{ev}}{2\varphi}(1 - (1 - t/\tau_{ev})^2) \quad . \quad (B13)$$

By setting $\Delta z = z_0$, the sedimentation time is obtained as

$$\tau_{sed}^{RH} = \tau_{ev}\left[1 - \left(1 - \frac{2\varphi z_0}{\tau_{ev} R_0^2}\right)^{1/2}\right] \quad . \quad (B14)$$

For $RH = 1$ no evaporation takes place and $\tau_{sed}^{RH} = \tau_{ev}$ is recovered. The distance by which the droplet falls during its evaporation time $\tau_{ev}$ follows as

$$\Delta z = \int_0^{\tau_{ev}} v(t) = \frac{R_0^2 \tau_{ev}}{2\varphi} = \frac{R_0^4}{2\theta\varphi(1-RH)} \quad . \quad (B15)$$

Equating the distance $\Delta z$ with the initial height $z_0$, the critical droplet radius follows from Eq. (B15) as

$$R_0^{crit} = (2\varphi\theta z_0(1-RH))^{1/4} \quad . \quad (B16)$$

Droplets with radii smaller than $R_0^{crit}$ will completely dry out before reaching the ground and the sedimentation time in Eq. (B14) diverges. Note that the calculations in this section neglect the finite solute concentration in the initial droplet, which will be considered in Appendix K and produces a lower limit to the droplet radius that can be obtained by evaporation.

**Appendix C: Stagnant droplet evaporation in the diffusion-limited regime with evaporation cooling effects**

There are several effects that temperature has on evaporation kinetics. The temperature diffusion constant is defined as $a = \lambda/(cC_P)$, where $\lambda$ is the heat conductivity coefficient, $c$ is the number density of the medium and $C_P$ is the molecular heat capacity of the medium at constant pressure. For air one finds a value $a_{air}$ = 2 x 10$^{-5}$ m²/s, which is very similar to the water diffusion constant in air (see Table I). Thus, temperature gradient effects cannot necessarily be neglected. The fact that evaporation cooling is relevant can be quickly appreciated. The molecular evaporation enthalpy of water at 25°C is $h_{ev} = 7.3 \times 10^{-20}$ J, the molecular heat capacity of liquid water at 25°C is $C_P^l = 1.3 \times 10^{-22}$ J, so one evaporating water molecules cools down 20 liquid water molecules from 25°C to 0°C. The molecular melting enthalpy of water is $h_m = 1.0 \times 10^{-20}$ J, so one evaporating water molecules freezes 7 liquid water molecules. Therefore, cooling due to evaporation needs to be accounted for.

The temperature profile around a spherical heat sink is described by the heat diffusion equation in radial coordinates

$$cC_P \frac{d}{dt} T(r,t) = \frac{\lambda_{air}}{r^2} \frac{d}{dr} r^2 \frac{d}{dr} T(r,t) \; , \tag{C1}$$

where $\lambda_{air}$ denotes the heat conductivity of air. The stationary temperature distribution is given by

$$T(r) = T_0 \left(1 - \frac{b_T}{r}\right) , \tag{C2}$$

where $T_0$ is the ambient temperature, from which the droplet surface temperature follows as

$$T_s = T_0 \left(1 - \frac{b_T}{R}\right) . \tag{C3}$$

The adiabatic approximation is used, meaning that the time it takes for the stationary temperature distribution to build up, is neglected; this is justified since the heat capacity of the droplet is small compared to the evaporation enthalpy, as shown above. After a few water molecules have evaporated, the droplet will have a uniform temperature equal to the air close to the surface. The heat flux into the droplet is given by

$$J_h = 4\pi R^2 \lambda_{air} \frac{d}{dR} T(R) = 4\pi \lambda_{air} T_0 b_T \; . \tag{C4}$$

In a stationary state, the thermal heat flux exactly balances the evaporation cooling rate, which is the water evaporation flux $J$ from Eq. (B5) times the evaporation enthalpy. The energy balance equation reads explicitly

$$J_h = h_{ev} J = 4\pi R^2 h_{ev} D_w \frac{k_e c_l - k_c c_0}{D_w + k_c R} = 4\pi R^2 h_{ev} D_w \frac{k_c c_g (1 - RH)}{D_w + k_c R} \; . \tag{C5}$$

In the diffusion-limited scenario this leads to

$$J_h = h_{ev} J = 4\pi R h_{ev} D_w c_g (1 - RH) \; . \tag{C6}$$

Combining Eqs. (C3), (C4) and (C6), one obtains for the temperature depression at the droplet surface

$$\Delta T = T_0 - T_s = \frac{D_w c_g h_{ev}}{\lambda_{air}} (1 - RH) = \varepsilon_T (1 - RH) \; , \tag{C7}$$

where the numerical prefactor for air is given by $\varepsilon_T \equiv \frac{D_w c_g h_{ev}}{\lambda_{air}} = 52$. This is a surprising result, as it would suggest that the evaporation of a droplet leads to droplet freezing at room temperature at all but very high relative humidities.

The estimate in Eq. (C7) neglects that there are counteracting effects that decrease the evaporation rate with decreasing temperature. Inspection of Table I and noting that the water vapor concentration is linearly proportional to the water vapor pressure, demonstrates that the dominant temperature effect in Eq. (C6) comes from the saturated water vapor concentration $c_g$, which is related to the liquid water density according to

$$c_g = c_l e^{\mu_{ex}/k_B T} \quad . \tag{C8}$$

The saturated water vapor concentration $c_g$ thus depends on the liquid water concentration $c_l$ (which depends only slightly on temperature) and exponentially on the water excess chemical potential, which at room temperature is given by $\mu_{ex} = -\frac{27 kJ}{mol}$ and decreases significantly with decreasing temperature, showing that $c_g$ quickly decreases as the temperature decreases. In the temperature range between 0°C and 25°C, the surface water vapor concentration can be linearly fitted according to

$$c_g^{surf} = c_g(1 - \varepsilon_c \Delta T) \quad , \tag{C9}$$

with a numerical prefactor $\varepsilon_C = 0.037 = (2340 - 610)/(20 \times 2340)$, which is obtained by linear interpolation of the water vapor pressure at 0°C, $P_v$ = 0.61 kPa, and at 20°C, $P_v$ = 2.34 kPa, see Table I. Replacing $c_g$ by $c_g^{surf}$ in Eq. (C6), the final result for the temperature depression at the droplet surface follows from solving the energy balance equation Eq. (C6) in a self-consistent manner, which gives

$$\Delta T = T_0 - T_s = \frac{\varepsilon_T(1-RH)}{1+\varepsilon_T \varepsilon_C} = 17(1 - RH). \tag{C10}$$

One sees that the temperature reduction at the droplet surface, including the temperature dependence of the water vapor concentration at the droplet surface, is less pronounced and for *RH* = 0.5 is about 9 K. Thus, freezing is preempted and the linearization of the water vapor concentration in Eq. (C9) is valid.

In the presence of evaporation-induced droplet cooling, the diffusional water flux is obtained from Eq. (C6) as

$$J = 4\pi R D_w(c_g^s - c_0) = 4\pi R D_w(c_g(1 - \varepsilon_C \Delta T) - c_0), \tag{C11}$$

which using Eq. (C10) can be rewritten as

$$J = 4\pi R D_w c_g \left(1 - \frac{\varepsilon_C \varepsilon_T}{1+\varepsilon_C \varepsilon_T}\right)(1 - RH) \,. \tag{C12}$$

Repeating the steps that led to Eq. (B9) in Appendix B, one obtains the modified evolution equation for the radius

$$R(t) = (1 - \theta\, t(1 - RH)/R_0^2)^{1/2}, \qquad (C13)$$

where the modified numerical prefactor is given by

$$\theta = 2 D_w c_g v_w \left(1 - \frac{\varepsilon_C \varepsilon_T}{1 + \varepsilon_C \varepsilon_T}\right) = 3.5 \times 10^{-10} m^2/s. \qquad (C14)$$

The factor that self-consistently accounts for the evaporation cooling effect is thus given by $\left(1 - \frac{\varepsilon_C \varepsilon_T}{1 + \varepsilon_C \varepsilon_T}\right) = 0.32$. Cooling considerably slows down the evaporation process but does not lead to droplet freezing.

**Appendix D: Stagnant droplet evaporation in the reaction-rate-limited regime**

To obtain the evaporation flux in the reaction-rate-limited regime, one starts from Eq. (B5) and assumes $k_c R < D_w$, which is realized for droplet radii $R < \frac{D_w}{k_c} = 100\ nm$ (see discussion after Eq. (B6)). In this limit, one can neglect the term proportional to $k_c R$ in the denominator of Eq. (B5) and obtain the reaction-rate-limited result for the flux, which is proportional to $R^2$ and thus to the surface area of the droplet,

$$J = 4\pi R^2 k_c c_g (1 - RH)\ . \qquad (D1)$$

Mass conservation of the droplet reads

$$\frac{d}{dt}\left(\frac{4\pi}{3} R^3(t)\right) = -v_w J = -4\pi R^2 k_c c_g v_w (1 - RH)\ . \qquad (D2)$$

The differential equation (D2) is easily solved with the result

$$R(t) = R_0 \left(1 - t\frac{k_c c_g v_w (1-RH)}{R_0}\right) = R_0(1 - \alpha^{reac} t(1-RH)/R_0)\ , \qquad (D3)$$

where $R_0$ is the initial droplet radius and the numerical prefactor is given by

$$\alpha^{reac} = k_c c_g v_w = 6.3 \times 10^{-3} m/s \qquad (D4)$$

and has units of a velocity. It is seen that shrinking of the radius now is linear in time, evaporation will stop at a radius when osmotic effects due to dissolved solutes and the presence of virions inside the droplet balance the evaporation chemical potential. The time needed to shrink the droplet to the final state can be approximated by the time needed to shrink the radius to zero, which from Eq. (D3) follows as

$$\tau_{ev}^{reac} = \frac{R_0}{\alpha^{reac}(1-RH)}\ . \qquad (D5)$$

For an initial radius of $R_0$ = 100nm, which is the upper limit where the stagnant reaction-limited evaporation regime is valid, the evaporation time is $\tau_{ev}^{reac} = 16\,\mu s/(1-RH)$. Thus, in completely dry air with $RH = 0$ the evaporation of a droplet with an initial radius of $R_0$ = 100 nm takes 16 $\mu s$; for relative humidity $RH$ = 0.5, the complete evaporation of a droplet with an initial radius of $R_0$ = 100 nm takes 32 $\mu s$. This time is completely negligible compared to the sedimentation time of a droplet with a radius of 100 nm, which is 20 days. The final reaction-rate limited evaporation stage can thus be neglected for droplets significantly larger than 100 nm in radius.

In fact, evaporation cooling effects can approximately be neglected in the reaction-rate-limited regime, since the heat transport stays diffusion limited and thus is faster than the reaction-rate-limited diffusive transport of water molecules.

**Appendix E: Convection effects on evaporation for laminar flow**

To estimate air convection effects on the evaporation rate at an analytically manageable level, the concept of a concentration boundary layer will be used. In this section laminar flow will be assumed, finite-Reynolds number effects will be considered in subsequent appendices. In order to obtain the concentration boundary layer thickness, it is necessary to calculate the time it takes for a water molecule to diffuse a certain distance away from the droplet surface. The water diffusion velocity is given by

$$u(r) = \frac{dr(t)}{dt} = \frac{j(r)}{c(r)} = \frac{-D_w dc(r)/dr}{c(r)} = \frac{-D_w b/r^2}{1+\frac{b}{r}-\frac{b}{R+\Delta}} \quad . \tag{E1}$$

In the last step the presence of a concentration boundary layer at a radius $r = R+\Delta$ was assumed, where the concentration boundary layer width is defined by $\Delta$ and the stationary density distribution is given by

$$c(r) = c_0\left(1 + \frac{b}{r} - \frac{b}{R+\Delta}\right) \quad . \tag{E2}$$

By this construction, at a radius $r = R+\Delta$, the water vapor concentration is equal to the ambient water vapor concentration $c_0$. Using again the reactive boundary condition at the sphere surface $r = R$, Eq. (B3), the coefficient b is given by

$$b = R^2 k_c \frac{c_g/c_0 - 1}{D_w + k_c \Delta R/(R+\Delta)} \cong R^2 k_c \frac{c_g/c_0 - 1}{D_w + k_c \Delta} \quad , \tag{E3}$$

where $k_e c_l = k_c c_g$ was used. One sees that for a concentration boundary layer larger than the droplet radius, i.e. for $\Delta > R$, the effect of $\Delta$ disappears. In the last step therefore the opposite limit, $\Delta < R$, was considered. The total water evaporation flux $J$ follows as

$$J = 4\pi R^2 j = 4\pi D_w c_0 b = 4\pi R^2 D_w k_c c_g \frac{1-RH}{D_w + k_c \Delta} \quad . \tag{E4}$$

The differential equation implied by Eq. (E1) is easily solved and gives

$$(r^3(t) - R^3)\left(1 - \frac{b}{R+\Delta}\right)/3 + b(r^2(t) - R^2)/2 = D_w bt \; . \tag{E5}$$

Using $r = R+\Delta$ and $\Delta < R$, this simplifies to

$$D_w b \tau_{diff} \approx \Delta R^2 \left(1 - \frac{b}{R+\Delta}\right) + b\Delta R \approx \Delta R^2 + b\Delta^2 ,  \quad (E6)$$

which defines the time it takes a water molecule to diffuse through the concentration boundary layer, which is called the diffusion time $\tau_{diff}$. Inserting the diffusion-limited approximation for $b$ from Eq. (E3), valid for $k_c \Delta > D_w$, the diffusion time becomes

$$\tau_{diff} = \frac{\Delta^2}{D_w(1-RH)} . \quad (E7)$$

This diffusion time has to be compared with the convection time scale

$$\tau_{conv} = \frac{\pi r/2}{v_{air}(r)} \approx \frac{\pi R^2}{2V\Delta} , \quad (E8)$$

which is the time the flow needs to travel a quarter circle around the droplet at the boundary layer located at a radius of $r = R+\Delta$ from the droplet center. In Eq. (E8) the air flow profile around the droplet for laminar flow conditions has been used, given approximately by $v_{air}(r) \approx V \frac{r-R}{r} \approx V\Delta/R$, where $V$ is the stationary droplet velocity from Eq. (A1b). By equating the diffusion and convection time scales $\tau_{diff}$ and $\tau_{conv}$ the concentration boundary layer thickness results as

$$\Delta = \Delta_0 (1 - RH)^{1/3} \quad (E9)$$

with

$$\Delta_0 = \left(\frac{9\pi\eta D_w}{4\rho g}\right)^{1/3} = 80 \; \mu m . \quad (E10)$$

Interestingly, the concentration boundary layer thickness does not depend on the speed with which the droplet falls to the ground, i.e., it does not depend on the droplet radius $R$. This result means that for radii larger than $R = 80$ μm and for dry air, one has $\Delta/R < 1$ and thus convection cannot be neglected and will accelerate evaporation.

The diffusion-limited rate scenario is defined by $k_c \Delta > D_w$, which using Eqs. (B6) and (E9), is satisfied for relative humidity $RH < 1\text{-}10^{-9}$, which is certainly always true. It transpires that convective evaporation in the laminar flow regime is basically always diffusion limited. Thus, the term proportional to $D_w$ in the denominator of Eq. (E4) can be safely neglected and the diffusion-limited flux is obtained

$$J = 4\pi R^2 D_w c_g (1 - RH)/\Delta . \quad (E11)$$

Mass conservation of the droplet can be written in terms of the droplet volume as

$$\frac{d}{dt}\left(\frac{4\pi}{3} R^3(t)\right) = -v_w J = -4\pi R^2 D_w c_g v_w (1 - RH)/\Delta . \quad (E12)$$

The differential equation (E12) is easily solved with the result

$$R(t) = R_0\left(1 - t\frac{D_w c_g v_w (1-RH)^{2/3}}{\Delta_0 R_0}\right) = R_0\left(1 - \alpha^{conv} t(1-RH)^{2/3}/R_0\right), \quad \text{(E13)}$$

where $R_0$ is the initial droplet radius and the numerical factor is given by

$$\alpha^{conv} = D_w c_g v_w / \Delta_0 = 7.9 \times 10^{-6} \, m/s \quad \text{(E14)}$$

and has units of a velocity. It is seen that shrinking of the radius is linear in time. Again, the evaporation time to a radius where osmotic effects due to dissolved solutes and the presence of virions inside the droplet balance the evaporation chemical potential can thus be approximated by the time needed to shrink the droplet radius down to zero, given by

$$\tau_{ev}^{conv} = \frac{R_0}{\alpha^{conv}(1-RH)^{2/3}}. \quad \text{(E15)}$$

Thus, at a relative humidity of $RH = 0.5$, a droplet with an initial radius of $R_0 = 1$ mm has an evaporation time of $\tau_{ev}^{conv} = 200$ s, but takes (neglecting the shrinking of the radius) $\tau_{sed} = 0.017$ s to fall to the ground from a height of 2 meters, so it will not dry out before being sedimenting to the ground. To accurately calculate the critical initial radius below which a droplet completely dries out before falling to the ground, Eq. (1) is rewritten in terms of the instantaneous droplet velocity and combined with Eq. (E13) to give

$$v(t) = \frac{D_R g m(t)}{k_B T} = \frac{R^2(t)}{\varphi} = \frac{R_0^2}{\varphi}\left(1 - \alpha^{conv} t(1-RH)^{\frac{2}{3}}/R_0\right)^2. \quad \text{(E16)}$$

The distance by which the droplet falls during its evaporation time, $\tau_{ev}^{conv}$, follows by integration as

$$\Delta z = \int_0^{\tau_{ev}^{conv}} v(t) = \frac{R_0^3}{3\alpha\varphi(1-RH)^{2/3}}. \quad \text{(E17)}$$

Now equating the distance $\Delta z$ with the height $z_0$, the critical droplet radius is

$$R^{crit} = \left(3\varphi\alpha^{conv} z_0 (1-RH)^{2/3}\right)^{1/3}, \quad \text{(E18)}$$

which for an initial height of $z_0 = 2$ m and for $RH = 0.5$ gives $R^{crit} = 67$ μm. This is about the same radius where convection effects are relevant from Eq. (E9). Thus, convective evaporation effects do not play a significant role for droplets that are released from a height of $z_0 = 2$ m. For an initial height of $z_0 = 2$ km it follows from Eq. (E18) that droplets with a radius smaller than 670 μm evaporate before they fall to the ground.

The results derived in this section neglect effects due to non-linear hydrodynamics, finite Reynolds number effects will be discussed in the following appendices. Evaporation cooling effects have not been treated explicitly in this appendix, but can be approximately accounted for multiplying the water vapor density $c_g$ in Eq. (E14) by the correction term $\left(1 - \frac{\varepsilon_C \varepsilon_T}{1+\varepsilon_C \varepsilon_T}\right)$ derived in Appendix C.

**Appendix F: Non-laminar flow effects and flow boundary layer**

Within the laminar flow boundary layer around an object, viscosity is relevant and laminar Stokes flow develops, outside this flow boundary layer potential flow is obtained. The flow boundary layer thickness scales as (36)

$$\delta = \left(\frac{vx}{V}\right)^{1/2} \tag{F1}$$

as a function of the distance *x* that the flow has moved along the object, where *V* is the velocity of the object and $v = \eta/\rho_{air}$ is the kinematic velocity. The density of air is denoted as $\rho_{air}$ and given in Table I. The kinematic viscosity has units of a diffusion constant and characterizes the diffusivity of momentum or vorticity, its value in air is $v = 1.5 \times 10^{-5}$ m²/s and thus it is half the value of the water diffusion constant in air. This shows that momentum diffuses slightly slower than water in air, thus the flow boundary layer $\delta$ is expected to be smaller than the concentration boundary layer $\Delta$. As result, flow boundary layer effects are expected to be relevant.

A simple estimate of the importance of momentum diffusion for a droplet moving in air is obtained by asking whether the flow boundary layer thickness $\delta$ is smaller than the droplet radius *R* for a flow that has travelled by a distance that corresponds to a quarter circle $x = \pi R/2$ around the droplet, i.e.

$$\delta = \left(\frac{v\pi R/2}{V}\right)^{1/2} < R \quad , \tag{F2}$$

which can be rewritten as

$$Re/\pi \equiv \frac{2\,RV}{\pi v} = \frac{2\,RV\rho_{air}}{\pi\eta} = \left(\frac{R}{R_*}\right)^3 > 1 \tag{F3}$$

and is equivalent to the condition that the Reynolds number *Re* is larger than $\pi$. The characteristic radius is defined as

$$R_* = \left(\frac{9\pi\eta^2}{4g\rho_{air}\rho R^3}\right)^{1/3} = 59\mu m \quad . \tag{F4}$$

The radius at which the Reynolds number becomes larger than $\pi$ and the flow around the sphere will not be laminar anymore is therefore given by $R_* = 59\mu m$. The typical momentum boundary layer thickness given in Eq. (F2) can be rewritten in terms of the characteristic radius as

$$\delta = \left(\frac{R_*^3}{R}\right)^{1/2} \quad , \tag{F5}$$

and is for a sphere with the characteristic radius given by $R_*$ itself. Since for not too humid air $R_*$ is smaller than the concentration boundary layer width $\Delta_0 = 80\,\mu m$ from Eq. (E10), it follows that convection effects are modified in the presence of a flow boundary layer, which will be treated in Appendices G and H.

**Appendix G: Convection effects on evaporation: Double boundary layer scenario in humid air**

In this section the double boundary-layer problem will be addressed, where both concentration and fluid flow boundary layers with widths $\Delta$ and $\delta$ are present. It will be assumed that $\Delta < \delta$, which a posteriori will be shown to correspond to humid air; the opposite case $\Delta < \delta$ for dry air will be treated in Appendix H.

Due to the presence of the flow boundary layer, the air flow field around the droplet is compressed by a factor $R/\delta$ and is approximately given by

$$v_{air}(r) \approx V \frac{R}{\delta} \left( \frac{r-R}{r} \right) \ . \tag{G1}$$

The convection time scale is defined as the time the flow needs to travel a quarter circle around the droplet at the concentration boundary layer located at a radius of $r = R+\Delta$ from the droplet center and is given by

$$\tau_{conv} = \frac{\pi r/2}{v_{air}(r)} \approx \frac{\pi R \delta}{2V\Delta} \ , \tag{G2}$$

which is smaller than the result for laminar flow in Eq. (E8) by a factor of $\delta/R$. By equating the diffusion time scale $\tau_{diff}$ in Eq. (E7), which is not modified by flow boundary-layer effects, with the convection time scale $\tau_{conv}$ in Eq. (G2), the boundary layer thickness results as

$$\Delta = \Delta_0 (1-RH)^{1/3} \left( \frac{R_*}{R} \right)^{1/2} \ , \tag{G3}$$

where $\Delta_0 = 80 \ \mu m$ is the laminar concentration boundary layer width from Eq. (E10) and $R_* = 59 \ \mu m$ is the characteristic radius for non-laminar flow effects from Eq. (F4). To check whether the assumption $\Delta < \delta$ used in this Appendix is satisfied, Eq. (G3) is divided by the momentum boundary layer width $\delta$ from Eq. (F5) to obtain

$$\frac{\Delta}{\delta} = (1-RH)^{1/3} \frac{\Delta_0}{R_*} \ . \tag{G4}$$

It transpires that the assumption $\Delta < \delta$ is satisfied for humid air with a relative humidity larger than

$$RH > 1 - \left( \frac{\Delta_0}{R_*} \right)^{-3} = 0.58 \ . \tag{G5}$$

When is the double-boundary-layer evaporation regime entered? This question is equivalent to asking when the boundary layer width $\Delta$ as given by Eq. (G3) becomes smaller than the radius $R$. For an intermediate humidity that coincides with the threshold value Eq. (G5), it is found that the double-boundary-layer evaporation regime is entered for radii $R > R_* =$

59μm, i.e., as soon as boundary flow effects occur. For more humid the threshold radius increases and follows from Eq. (G3).

The diffusion-limited rate scenario is defined by $k_c\Delta > D_w$, which using Eqs. (B6) and (G3), is satisfied for radii

$$R < (1 - RH)^{2/3} \left(\frac{\Delta_0^2 R_*}{(D_w/k_c)^2}\right) = (1 - RH)^{2/3}\, 38\, m \ . \tag{G6}$$

It transpires that double-boundary layer convective evaporation in humid air is always diffusion limited. Thus, the differential equation Eq. (E12) that is valid in the diffusion limit can be used, which in conjunction with Eq. (G3) yields

$$R(t) = \left(R_0^{1/2} - t\frac{D_w c_g v_w (1-RH)^{2/3}}{2\Delta_0 R_*^{1/2}}\right)^2 = \left(R_0^{1/2} - \alpha^{dbl1} t (1-RH)^{2/3}\right)^2 , \tag{G7}$$

where $R_0$ is the initial droplet radius and the numerical factor is given by

$$\alpha^{dbl1} = \frac{D_w c_g v_w}{2\Delta_0 R_*^{1/2}} = 5.7 \times 10^{-4} m^{1/2}/s \ . \tag{G8}$$

It is seen that the shrinking the radius slows down over time. Evaporation cooling effects have not been treated explicitly, but can be approximately accounted for by multiplying the water vapor density by the correction term $\left(1 - \frac{\varepsilon_C \varepsilon_T}{1+\varepsilon_C \varepsilon_T}\right)$ derived in Appendix C.

**Appendix H: Convection effects on evaporation: Double boundary layer scenario in dry air**

In this section, it will be assumed that the concentration boundary layer width is larger than the flow boundary layer width, i.e. $\Delta > \delta$, which a posteriori will be shown to correspond to dry air. The calculation closely follows Appendix G.

At the concentration boundary layer, since $\Delta > \delta$, the air flow field around the droplet is unperturbed by the presence of the droplet and given by $v_{air}(r) \approx V$. The convection time scale is therefore given by

$$\tau_{conv} = \frac{\pi r/2}{v_{air}(r)} \approx \frac{\pi R}{2V} , \tag{H1}$$

which is smaller than the corresponding result in Eq. (G2). By equating the diffusion time scale $\tau_{diff}$ in Eq. (E7), which is not modified by flow boundary-layer effects, with the convection time scale $\tau_{conv}$ in Eq. (H1), the boundary layer thickness results as

$$\Delta = (1 - RH)^{1/2} \left(\frac{\Delta_0^3}{R}\right)^{1/2} , \tag{H2}$$

where $\Delta_0 = 80\ \mu m$ is the laminar concentration boundary layer width from Eq. (E10). To check when the assumption $\Delta > \delta$ used here is satisfied, Eq. (H2) is divided by the momentum boundary layer width $\delta$ from Eq. (F5) to obtain

$$\frac{\Delta}{\delta} = (1 - RH)^{1/2} \left(\frac{\Delta_0}{R_*}\right)^{3/2} \quad . \tag{H3}$$

It transpires that the assumption $\Delta > \delta$ is satisfied for dry air with a relative humidity smaller than

$$RH < 1 - \left(\frac{\Delta_0}{R_*}\right)^{-3} = 0.58 \quad . \tag{H4}$$

When is the double-boundary-layer evaporation regime entered? This question is equivalent to asking when the boundary layer width $\Delta$ as given by Eq. (H2) becomes smaller than the radius $R$. For an intermediate humidity that coincides with the threshold value Eq. (H4), it follows that the double-boundary-layer evaporation regime is entered for radii $R > R_* = 59\mu m$, i.e., as soon as boundary flow effects according to Eq. (F4) occur.

The diffusion-limited rate scenario is defined by $k_c \Delta > D_w$, which using Eqs. (B6) and (H2), is satisfied for radii

$$R < (1 - RH)\left(\frac{\Delta_0^3}{(D_w/k_c)^2}\right) = (1 - RH)\, 51\, m \quad . \tag{H5}$$

It transpires that double-boundary layer convective evaporation in dry air is always diffusion limited. Thus, the differential equation Eq. (E12) can be used, which in conjunction with Eq. (H2) yields

$$R(t) = \left(R_0^{1/2} - t\frac{D_w c_g v_w (1-RH)^{1/2}}{2\Delta_0^{3/2}}\right)^2 = \left(R_0^{1/2} - \alpha^{dbl2} t(1-RH)^{1/2}\right)^2 \quad , \tag{H6}$$

where $R_0$ is the initial droplet radius and the numerical factor is given by

$$\alpha^{dbl2} = \frac{D_w c_g v_w}{2\Delta_0^{3/2}} = 3.5 \times 10^{-4} m^{1/2}/s \quad . \tag{H7}$$

Evaporation cooling effects have not been treated explicitly, but can be approximately accounted for by multiplying the water vapor density $c_g$ by the correction term $\left(1 - \frac{\varepsilon_C \varepsilon_T}{1+\varepsilon_C \varepsilon_T}\right)$ derived in Appendix C.

In summary, for droplet radii larger than about $R = 59$ μm, non-linear hydrodynamic effects become important and produce a finite, so-called flow boundary layer, around the falling droplet. Inside the flow boundary layer viscous effects are relevant and laminar flow is obtained, outside the flow boundary layer viscous effects can be neglected and potential flow is realized. At about the same range of radii, the stagnant approximation becomes invalid, because convection speeds up the evaporation process. This effect can be described by a concentration boundary layer. The problem is thus a double-boundary-layer problem and involves a concentration and a flow boundary layer. Whether the concentration boundary layer or the flow boundary layer is smaller and thus more relevant, depends on the relative air humidity. It turns out that the evaporation in the presence of convection is diffusion limited. For humid air with a relative humidity $RH > 0.59$, the concentration boundary layer is evaporation-rate limiting and the time-dependent radius decrease is given by Eq. (G7). For drier air with a relative humidity $RH < 0.59$, the flow boundary layer is evaporation-rate limiting and the time dependent radius decrease is given by Eq. (H6). The results for the humid and dry boundary layer scenarios thus look quite similar, but the physical mechanisms behind the evaporation process are very different. At high Reynolds numbers the friction experienced by a falling droplet is reduced due to a combination of boundary layer effects, boundary-layer separation effects and turbulence effects. In Appendix I it is shown that the Stokes expression for the friction force acting on a falling spherical droplet is accurate for radii below about 160 µm.

**Appendix I: Falling speed for large Reynold numbers**

The Stokes approximation used for calculating the stationary falling speed of droplets in Eq. (A1b) is a low-Reynolds number approximation. An empirical formula for the settling velocity of a sphere in air, that is valid over the entire range of Reynold numbers, is (36)

$$V = \sqrt{\frac{8gR\rho}{3\rho_{air}c_D}} \quad , \tag{I1}$$

where the resistance coefficient is given by

$$c_D = \frac{24}{Re} + \frac{4}{Re^{1/2}} + 0.4 \tag{I2}$$

and the Reynolds number $Re$ is defined in Eq. (F3). The result of Eq. (A1b) is reproduced by Eq. (I1) when only the first term in Eq. (I2) is used. The accuracy of this low-Reynolds number approximation can be checked by comparing the first and third terms in Eq. (I2), which become equal for a Reynolds number of about $Re=60$, which corresponds, using again Eq. (F3), to a radius of about $(\frac{60}{\pi})^{1/3} R_* = 157$ µm. This suggests that the falling speed according to Eq. (A1b) is quite accurate for radii below 157 µm. For larger radii the falling speed will be reduced and thus the sedimentation time will be increased. Therefore, the sedimentation times presented in this note are lower estimates for radii larger than about 157 µm.

**Appendix J: Internal mixing effects**

The calculations so far assumed that diffusion inside the droplet is sufficiently rapid, so that the water concentration at the droplet surface does not differ significantly from the mean water concentration in the droplet. It will turn out that this is a limiting factor for the maximal droplet size that can evaporate at the speed predicted here. According to the diffusion law, the time it takes for a water molecule to diffuse over the droplet radius $R$ inside the droplet is

$$\tau_{mix} = \frac{R^2}{2D_w^l} \quad , \tag{J1}$$

where $D_w^l$ is the molecular water self-diffusion constant in liquid water. The mixing time within a droplet of radius $R$ = 10 µm is $\tau_{mix}$ = 25 ms and inside a droplet of radius $R$ = 100 µm it is $\tau_{mix}$ =2.5 s. Equating $\tau_{mix}$ with the mean sedimentation time Eq. (1), the mixing time inside the droplet is only shorter than the sedimentation time for radii smaller than about 100 µm. For larger droplets the internal diffusion will slow down evaporation. Convection effects inside the droplet, due to shear coupling to the air outside flow field, will counteract this effect, but are not considered here. Also, the increase of the internal droplet viscosity due to increasing solute concentration and the possible presence of a solid solute phase is not considered and will further slowdown the diffusion inside droplets.

**Appendix K: Solute-induced vapor pressure reduction effects**

Any solute present in the aqueous droplet decreases the water vapor pressure. This colligative effect is basically due to the dilution of the liquid water and can be derived in the following fashion:

*Water chemical potential in a two-component liquid system:*

The entropy of a liquid two-component system consisting of $N_w$ water molecules with molecular volume $v_w$ and $N_s$ solutes molecules with molecular volume $v_s$ is given up to an irrelevant constant by

$$\frac{S}{k_B} = -N_w \ln\left(\frac{N_w}{N_w v_w + N_s v_s}\right) - N_s \ln\left(\frac{N_s}{N_w v_w + N_s v_s}\right), \quad (K1)$$

where ideal mixing and ideal volume additivity is assumed. The water chemical potential in the liquid follows as

$$\frac{\mu_w^l}{k_B T} = -\frac{\partial S}{k_B \partial N_w} + \frac{\mu_{ex}}{k_B T} = \ln\left(\frac{1-\Phi}{v_w}\right) + \Phi\left(1 - \frac{v_w}{v_s}\right) + \frac{\mu_{ex}}{k_B T}, \quad (K2)$$

where $\Phi = N_s v_s / (N_w v_w + N_s v_s)$ is the solute volume fraction. In the limit $\Phi \to 0$ this can be rewritten as

$$\frac{\mu_w^l}{k_B T} \approx \ln\left(\frac{1-\Phi v_w/v_s}{v_w}\right) + \frac{\mu_{ex}}{k_B T}. \quad (K3)$$

When the solute volume fraction is finite and in particular when the water and solute molecular volumes are similar to each other, one can instead rewrite Eq. (K2) as

$$\frac{\mu_w^l}{k_B T} \approx \ln\left(\frac{1-\Phi}{v_w}\right) + \frac{\mu_{ex}}{k_B T}. \quad (K4)$$

*Water chemical potential in a multi-component liquid system:*

The entropy of a liquid many-component system consisting of $N_w$ water molecules with molecular volume $v_w$ and $N_i$ solute molecules of type *i* with molecular volume $v_i$ each, where *i=1 … M*, is given up to an irrelevant constant by

$$\frac{S}{k_B} = -N_w \ln\left(\frac{N_w}{N_w v_w + \sum_i v_i N_i}\right) - \sum_i N_i \ln\left(\frac{N_i}{N_w v_w + \sum_j v_j N_j}\right), \quad (K5)$$

where again ideal mixing and ideal volume additivity was assumed. The water chemical potential in the liquid follows as

$$\frac{\mu_w^l}{k_B T} = -\frac{\partial S}{k_B \partial N_w} + \frac{\mu_{ex}}{k_B T} = \ln\left(\frac{1-\Phi}{v_w}\right) + \sum_i \Phi_i \left(1 - \frac{v_w}{v_i}\right) + \frac{\mu_{ex}}{k_B T}, \quad (K6)$$

where $\Phi_i = N_i v_i / (N_w v_w + \sum_i v_i N_i)$ is the solute volume fraction of species i and $\sum_i \Phi_i = \Phi$. In the limit $\Phi \to 0$ this can be written as

$$\frac{\mu_w^l}{k_B T} \approx \ln\left(\frac{1 - \sum_i \Phi_i v_w / v_i}{v_w}\right) + \frac{\mu_{ex}}{k_B T}. \quad (K7)$$

On the other hand, when the solute volume fraction $\Phi$ is finite and the sum $\sum_i \Phi_i \left(1 - \frac{v_w}{v_i}\right)$ is small, one can instead write

$$\frac{\mu_w^l}{k_B T} \approx \ln\left(\frac{1-\Phi}{v_w}\right) + \frac{\mu_{ex}}{k_B T}, \quad (K8)$$

which is the approximation that will be used in the following.

*Water vapor concentration and evaporation rate in presence of solutes:*

From the ideal expression for the water vapor chemical potential

$$\frac{\mu_w^g}{k_B T} = \ln(c_g) \quad (K9)$$

and the equality of chemical potentials, $\mu_w^g = \mu_w^l$, the equilibrium vapor concentration in the presence of solutes follows from Eq. (K8) as

$$c_g^{sol} = \left(\frac{1-\Phi}{v_w}\right) e^{\mu_{ex}/k_B T} = (1-\Phi) c_l e^{\mu_{ex}/k_B T}, \quad (K10)$$

which depends exponentially on the water excess chemical potential, at room temperature given by $\mu_{ex} = -\frac{27 kJ}{mol}$, and where the liquid water concentration in the absence of solute is denoted as $c_l = 1/v_w$. Assuming that initially the volume fraction of solutes is $\Phi_0$ and the initial radius is $R_0$, the water concentration in the liquid droplet with reduced radius R follows as

$$c_l^{sol} = \frac{1}{v_w}\left(1 - \Phi_0 \frac{R_0^3}{R^3}\right). \quad (K11)$$

Similarly, one obtains for the water vapor concentration in the presence of solutes

$$c_g^{sol} = c_g \left(1 - \Phi_0 \frac{R_0^3}{R^3}\right). \quad (K12)$$

Here, $c_g$ represents the water vapor concentration in the absence of solutes. Non-ideal effects can be included via the excess chemical potential and would be described by an activity coefficient different from unity, which is not pursued here. Replacing $c_g$ by $c_g^{sol}$ in Eq. (C12), one arrives at

$$J = 4\pi R D_w \left(1 - \frac{\varepsilon_C \varepsilon_T}{1+\varepsilon_C \varepsilon_T}\right)\left(c_g^{sol} - c_0\right). \tag{K13}$$

Together with Eq. (K12), one obtains the modified diffusive water flux in the presence of solutes as

$$J = 4\pi R D_w \left(1 - \frac{\varepsilon_C \varepsilon_T}{1+\varepsilon_C \varepsilon_T}\right)\left(c_g\left(1 - \Phi_0 \frac{R_0^3}{R^3}\right) - c_0\right). \tag{K14}$$

The mass conservation equation follows as

$$\frac{d}{dt}\left(\frac{4\pi}{3}R^3(t)\right) = -v_w J = -2\pi R \theta \left(1 - \Phi_0 \frac{R_0^3}{R^3} - RH\right), \tag{K15}$$

where $\theta$ is defined in Eq. (C14). Equation (K15) gives rise to the differential equation

$$\frac{2RdR}{1-R_{ev}^3/R^3} = -\theta(1-RH)dt, \tag{K16}$$

where the equilibrium droplet radius that is obtained in the long-time limit is defined as

$$R_{ev} = R_0 \left(\frac{\Phi_0}{1-RH}\right)^{1/3}. \tag{K17}$$

Here, $R_0$ is the initial radius and $\Phi_0$ is the initial volume fraction of solutes, including strongly bound hydration water. Only for RH = 0 does a droplet dry out to the minimal possible radius of $R_{ev} = R_0(\Phi_0)^{1/3}$, for finite relative humidity the equilibrium droplet radius is characterized by an equilibrium solute volume fraction of $\Phi_{ev} = 1 - RH$. As an example, for RH = 0.5, the free water and solute (including hydration water) volume fractions in the equilibrium state equal each other. Equation (K17) is modified for solutes that perturb the water activity, but for most solutes non-ideal water solution effects can be neglected. To illustrate this: the saturation concentration of NaCl in water is approximately 6 M, which corresponds, for simplicity assuming equal volume of Na$^+$ cations, Cl$^-$ anions and water molecules, roughly to a volume fraction of $\Phi_\infty$ =12M/(55+12)M = 0.18, thus suggesting a water vapor pressure corresponding to a humidity of $RH = 1 - \Phi_\infty = 0.82$, which is rather close to the experimental humidity created by a saturated NaCl solution of 0.75. This reflects that the activity coefficient of NaCl is rather close to unity for concentrations close to the solubility limit.

The solution of the differential equation (K16) can be written as

$$t(R) = \frac{R_{ev}^2}{\theta(1-RH)}\left[\mathcal{L}\left(\frac{R_0}{R_{ev}}\right) - \mathcal{L}\left(\frac{R}{R_{ev}}\right)\right], \tag{K18}$$

where the scaling function is given by

$$\mathcal{L}(x) = x^2 - \frac{2}{\sqrt{3}}\arctan\left(\frac{1+\frac{2}{x}}{\sqrt{3}}\right) - \frac{1}{3}\ln\left(\frac{x^2+x+1}{(x-1)^2}\right). \tag{K19}$$

The scaling function exhibits the asymptotic behavior

$$\mathcal{L}(x) \cong x^2 \qquad (K20)$$

for large arguments and

$$\mathcal{L}(x) \cong +\frac{2}{3}\ln(1-1/x) \qquad (K21)$$

for small arguments $x \to 1$. A quite accurate crossover expression is produced by summing the two limits as

$$\mathcal{L}(x) \cong x^2 + \frac{2}{3}\ln(1-1/x) \; . \qquad (K22)$$

Neglecting the logarithmic term in Eq. (K22) that reflects the kinetic slowing down due to the reduced water vapor pressure, one obtains from Eq. (K18) the limiting result

$$t(R)/\tau_{ev} = 1 - \frac{R^2}{R_0^2} \; , \qquad (K23)$$

from which an approximate expression for the evaporation time in the presence of solutes follows as

$$\tau_{ev}^{sol} = \tau_{ev}\left(1 - \frac{R_{ev}^2}{R_0^2}\right) \; . \qquad (K24)$$

The threshold radius below which the presence of solutes becomes important, can be defined by the radius where the function $t(R)$ changes curvature. The second derivative of $t(R)$ in Eq. (12) is given by

$$\frac{R_0^2 d^2 t(R)/\tau_{ev}}{dR^2} = -2 + \frac{2}{3}\left((1 - R/R_{ev})^{-2} - (R/R_{ev})^{-2}\right) \; ,$$

and vanishes at a radius of $R/R_{ev} = 1.54$. Thus, according to this curvature criterion, droplets enter the solute-dominated evaporation regime for radii smaller than $R = 1.54\, R_{ev}$, independent of the initial droplet radius $R_0$.

*Sedimentation time in the presence of solutes:*

The sedimentation of small enough droplets proceeds in two stages: First, the droplets shrink down to a radius given by Eq. (K17), second, the droplets sediment for an extended time with a fixed radius. The distance by which the droplet falls during its evaporation time $\tau_{ev}^{sol}$ follows in analogy to Eq. (B15) as

$$z_{ev} = \int_0^{\tau_{ev}^{sol}} v(t) = \frac{R_0^2 \tau_{ev}}{2\varphi}\left(1 - (1 - \tau_{ev}^{sol}/\tau_{ev})^2\right) = \frac{R_0^2 \tau_{ev}}{2\varphi}\left(1 - \left(\frac{R_{ev}}{R_0}\right)^4\right) \; . \qquad (K25)$$

Thus, the total sedimentation time is given by

$$\tau_{sed}^{sol} = \tau_{ev}^{sol} + \frac{\varphi(z_0 - z_{ev})}{R_{ev}^2} \quad , \tag{K26}$$

where the first term is the time it takes for the droplets to shrink down to the equilibrium radius $R_{ev}$ and the second term is the time it takes to sediment from the height $z_0 - z_{ev}$ to the ground as given by Eq. (1). Using Eq. (K24) and (K25) the final result for the sedimentation time can be written as

$$\tau_{sed}^{sol} = \frac{\varphi z_0}{R_{ev}^2} - \frac{\tau_{ev}}{2}\left(\frac{R_0}{R_{ev}} - \frac{R_{ev}}{R_0}\right)^2 . \tag{K27}$$

For droplets that are so large that they do not reach the radius $R_{ev}$ before they hit the ground, Eq. (B14) describes the sedimentation time accurately. The crossover between the two sedimentation time regimes occurs when $z_{ev}$ as described by Eq. (K25) equals $z_0$, the critical droplet radius follows as

$$R_0^{crit} = \left(\frac{2\varphi\theta z_0(1-RH)}{1-(\Phi_0/(1-RH))^{4/3}}\right)^{1/4} , \tag{K28}$$

which constitutes a generalization of Eq. (B16) in the presence of solutes. Droplets with radii smaller than $R_0^{crit}$ will reach their equilibrium radius before sedimenting to the ground and the sedimentation time is given by Eq. (K27).

**Appendix L: Surface tension effects**

The large surface tension of water increases the vapor pressure produced by droplets. The surface free energy of a droplet is given by

$$F = 4\pi\gamma R^2 . \tag{L1}$$

The chemical potential contribution, the so-called Kelvin potential, reads

$$\mu_{Kel} = \frac{dF}{dN} = \frac{dR}{dN}\frac{dF}{dR} = \frac{2\gamma v_w}{R} , \tag{L2}$$

where the number of water molecules inside the droplet is taken as $N = 4\pi R^3/(3v_w)$. Inserting numbers, the rescaled Kelvin potential reads

$$\frac{\mu_{Kel}}{k_B T} = \frac{10^{-9} m}{R} , \tag{L3}$$

which is significant compared to the water excess chemical potential only for droplet radii smaller than one nanometer.
To avoid confusion: the Laplace pressure

$$P_{Lap} = -\frac{dF}{dV} = -\frac{dR}{dV}\frac{dF}{dR} = \frac{2\gamma}{R} \tag{L4}$$

is significant and reaches 1 bar for a droplet radius of 1 μm, but it is unrelated to the vapor pressure.